# Structural and Electronic Properties of a Triangular Lattice Magnet NaPrTe$_2$ Compared with NaNdTe$_2$ and NaTbTe$_2$

Koki Eto[1], Yoshihiko Okamoto[1,*,†], Naoyuki Katayama[1], Hajime Ishikawa[2],
Koichi Kindo[2], and Koshi Takenaka[1]

[1]*Department of Applied Physics, Nagoya University, Nagoya 464-8603, Japan*
[2]*Institute for Solid State Physics, University of Tokyo, Kashiwa 277-8581, Japan*

NaPrTe$_2$, NaNdTe$_2$, and NaTbTe$_2$ are found to be triangular lattice magnets with the α-NaFeO$_2$ structure, where lanthanoid atoms with 4$f$ electrons form a triangular lattice, based on the structural analysis and physical property measurements of synthesized polycrystalline samples. The α-NaFeO$_2$ structure is a new polymorph of NaPrTe$_2$, which has been reported to crystallize in the cubic LiTiO$_2$ structure. Polytypism in NaPrTe$_2$ was discussed based on the structural parameters determined by the Rietveld analysis. NaPrTe$_2$ is suggested to be in the proximity of the phase boundary between the LiTiO$_2$ and α-NaFeO$_2$ types, as compared to NaNdTe$_2$ and NaTbTe$_2$, indicating that this compound might be interesting from the perspectives of the dimensional control of geometrically frustrated lattices. The magnetic susceptibility and heat capacity data indicated that NaPrTe$_2$ do not show long-range magnetic order or a spin-glass transition above 2 K.

## 1. Introduction

Many compounds with the chemical formula NaLnX$_2$, where Ln and X are lanthanoid and group 16 elements, respectively, have been synthesized so far. Most of them have NaCl-based crystal structures. When X is sulfur or selenium, almost all compounds crystallize in the disordered NaCl or α-NaFeO$_2$ structure.[1-10] In the former case, the cation sites of the NaCl structure are randomly occupied by Na and Ln atoms. In contrast, the α-NaFeO$_2$ structure is an ordered-NaCl-type structure with rhombohedral $R\bar{3}m$ symmetry, where the cation sites are alternately occupied by Na and Ln atoms, each of which form a two-dimensional triangular lattice [Fig. 1(a)]. NaLnS$_2$ and NaLnSe$_2$ with light lanthanoid Ln$^{3+}$ ions having ionic radii comparable to that of Na$^+$ tend to crystallize in the disordered NaCl structure, while those with the heavy Ln$^{3+}$ ions with smaller ionic radii favor the α-NaFeO$_2$ structure.[5,11] When X is oxygen, various ordered-NaCl-type structures, including α-NaFeO$_2$, can be realized.[11-13]

In recent years, α-NaFeO$_2$-type NaLnX$_2$ (X = O, S, and Se) has attracted attention from the perspective of geometrically frustrated magnetism due to the 4$f$ electrons of the lanthanoid atoms. A typical example is NaYbO$_2$,[14-16] where the Yb$^{3+}$ ions with spins of effective total angular momentum of $J_{\mathrm{eff}}$ = 1/2 form a triangular lattice. Although there is an antiferromagnetic interaction between the localized spins, this compound does not show magnetic order down to 50 mK.[14] Furthermore, it is noteworthy that the up-up-down magnetic order appears when applying a magnetic field of several T at a temperature of 1 K or lower.[14,15] This result suggests that the ground state of this material at a zero magnetic field is not a simple spin-disordered or spin glass state but instead a quantum spin liquid state realized by geometrical frustration of the triangular lattice. $NaYbS_2$ and $NaYbSe_2$ also do not show magnetic order down to the lowest measured temperature.[17-19] Although 4$f$ electron systems with a triangular lattice have not been shown to exhibit geometrically frustrated magnetism compared to $d$-electron systems and organic compounds, these results indicate that the $NaLnX_2$ family is promising as frustrated spin systems.

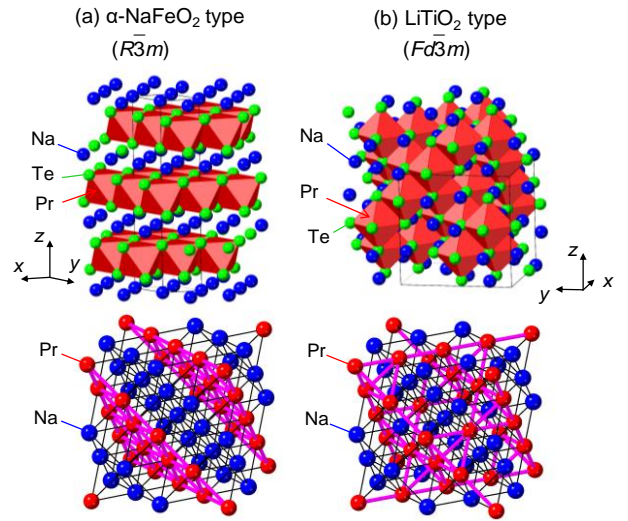

Figure 1. Crystal structures of (a) rhombohedral α-NaFeO$_2$ and (b) cubic LiTiO$_2$-type NaPrTe$_2$. For (a) and (b), the crystallographic parameters shown in Table I and Ref. 20 are used, respectively. In the upper panel, Pr atoms occupy the center of octahedra and the solid line indicates the unit cell. The lower panel shows the Na and Pr sublattices so as to contrast the arrangement of Na and Pr atoms in the α-NaFeO$_2$ and LiTiO$_2$ structures. The thick solid lines in the lower panel of (a) and (b) show the triangular lattice and pyrochlore structure made of Pr$^{3+}$ atoms, respectively.

In contrast to oxides, sulfides, and selenides, the $NaLnTe_2$ compounds have not been synthesized thus far, with the exception of $NaPrTe_2$. $NaPrTe_2$ was obtained as a by-product of $Pr_4N_2Te_3$ in a single crystalline form and reported to have the cubic $LiTiO_2$ structure shown in Fig. 1(b).[20] This crystal structure is one of the ordered NaCl structure and has the $Fd\text{-}3m$ space group. Each of Li and Ti atoms form a three-dimensional pyrochlore structure, where regular tetrahedra share their vertices. In this study, $NaPrTe_2$, $NaNdTe_2$, and $NaTbTe_2$ are found to be triangular lattice magnets with the α-$NaFeO_2$ structure, where $Ln^{3+}$ atoms carry localized spins. Moreover, $NaPrTe_2$ is found to be a polytypic compound, which can lead to the dimensional control of geometrically frustrated lattice.

## 2. Experiment

Polycrystalline samples of $NaLnTe_2$ (Ln = Pr, Nd, and Tb) were prepared by the solid-state reaction method. A stoichiometric amount of $Na_2Te$ powder (Kojundo Chemical Lab., 99%), Ln chips (Rare Metallic, 99.9%), and Te powder (Rare Metallic, 99.99%) was mixed in a glove box in an inert atmosphere. The mixture was put in an alumina crucible, which was sealed in an evacuated quartz tube. The tube was heated to and maintained at 873 K for 24 h, 1173 K (Ln = Pr and Tb) or 1273 K (Ln = Nd) for 12 h and then furnace cooled to room temperature. The obtained samples were decomposed in the air within a few hours, so we handled them in an inert gas atmosphere. The crystal structures of $NaLnTe_2$ were determined by Rietveld analysis using the RIETAN-FP program for the powder X-ray diffraction (XRD) patterns obtained at room temperature by employing synchrotron X-rays at BL02B2 of SPring-8 and at BL5S2 of Aichi Synchrotron Radiation Center.[21] The wavelengths of the X-rays used were λ = 0.620143, 0.61992, and 0.689015 Å for Ln = Pr, Nd, and Tb, respectively.

The magnetic susceptibility between 2 and 300 K was measured in a Magnetic Property Measurement System (Quantum Design). Magnetization measurements up to 61 T were performed using a multilayered pulse magnet with a duration of 4 ms. The magnetizations were measured at 1.4 K using the electromagnetic induction method employing a coaxial pick-up coil. The heat capacity above 2 K was meas-ured by the relaxation method in a Physical Property Measurement System (Quantum Design). The electrical resistivity above 2 K was measured by the four-probe method in a Physical Property Measurement System (Quantum Design).

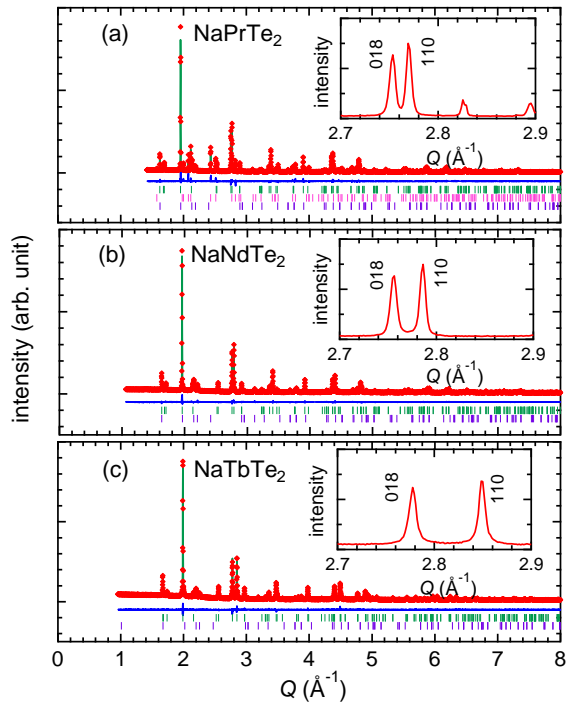

Figure 2. Synchrotron powder XRD patterns of (a) $NaPrTe_2$, (b) $NaNdTe_2$, and (c) $NaTbTe_2$ polycrystalline samples taken at room temperature. Filled circles show experimental data. The overplotted curve shows the calculated pattern, while the lower curve shows a difference plot between the experimental and calculated intensities. The upper and lower vertical bars indicate the position of the Bragg reflections of the main and $Ln_2O_2Te$ impurity phases, respectively. The middle bars in (a) indicate those of the $PrTe_2$ impurity phase. Each inset shows the XRD pattern around the 018 and 110 diffraction peaks based on the hexagonal axes.

## 3. Results and Discussion
### 3.1 Structural Properties

Figure 2 shows the synchrotron XRD patterns of the $NaLnTe_2$ (Ln = Pr, Nd, and Tb) polycrystalline samples taken at room temperature and the results of their Rietveld analysis. The crystallographic parameters determined by the analysis are listed in Table I. There were some small peaks caused by $Pr_2O_2Te$ and $PrTe_2$, $Nd_2O_2Te$, and $Tb_2O_2Te$ impurities for Ln = Pr, Nd, and Tb, respectively. Therefore, we performed the multiphase analysis including these impurity phases. In all cases, the differences between the experimental and calculated intensities were minimized when the rhombohedral α-$NaFeO_2$-type structural model was used as the main phase, indicating that the α-$NaFeO_2$-type $NaLnTe_2$ was obtained. The volume fractions of Pr$_2$O$_2$Te and PrTe$_2$ impurities in the NaPrTe$_2$ sample, Nd$_2$O$_2$Te impurity in the NaNdTe$_2$ sample, and Tb$_2$O$_2$Te impurity in the NaTbTe$_2$ sample estimated by the Rietveld analysis are 6%, 15%, 13%, and 7%, respectively.

Table I. Crystallographic parameters for NaPrTe$_2$, NaNdTe$_2$, and NaTbTe$_2$ determined by Rietveld analysis of synchrotron powder XRD data. The reliability factors of the refinements are also shown. The space group is $R\text{–}3m$ and the crystallographic parameters are shown using the hexagonal axes. The lattice parameters are $a$ = 4.53583(3), 4.51043(3), and 4.41051(4) Å and $c$ = 22.42581(17), 22.4595(2), and 22.4556(3) Å for NaPrTe$_2$, NaNdTe$_2$, and NaTbTe$_2$, respectively. The thermal displacement parameter $B$ for Na in NaPrTe$_2$ is constrained to 1.58, which is the refined $B$ value of Na in NaNdTe$_2$, because an inappropriate value was obtained in the refinement process probably due to the smaller atomic number of Na compared to Pr and Te.

|    |    | $x$ | $y$ | $z$ | $g$ | $B$ (Å$^2$) |
|----|----|-----|-----|-----|-----|-------------|
| NaPrTe$_2$ ($R_{wp}$ = 6.077%, $R_p$ = 5.447%, $R_e$ = 4.649%, $S$ = 1.3070) | | | | | | |
| Na | 3a | 0 | 0 | 0 | 1 | 1.58 |
| Pr | 3b | 0 | 0 | 1/2 | 1 | 0.91(2) |
| Te | 6c | 0 | 0 | 0.24659(4) | 1 | 0.942(18) |
| NaNdTe$_2$ ($R_{wp}$ = 5.943%, $R_p$ = 4.869%, $R_e$ = 5.139%, $S$ = 1.1565) | | | | | | |
| Na | 3a | 0 | 0 | 0 | 1 | 1.58(19) |
| Nd | 3b | 0 | 0 | 1/2 | 1 | 0.93(3) |
| Te | 6c | 0 | 0 | 0.24622(5) | 1 | 0.94(2) |
| NaTbTe$_2$ ($R_{wp}$ = 4.688%, $R_p$ = 3.814%, $R_e$ = 3.861%, $S$ = 1.2143) | | | | | | |
| Na | 3a | 0 | 0 | 0 | 1 | 1.1(3) |
| Tb | 3b | 0 | 0 | 1/2 | 1 | 0.87(4) |
| Te | 6c | 0 | 0 | 0.24468(6) | 1 | 0.89(4) |

It can be clearly seen from the inset figures in Fig. 2 that the crystal structures of the obtained NaLnTe$_2$ samples are not cubic, such as in the disordered-NaCl and LiTiO$_2$ structures, but instead have rhombohedral distortion. The 018 and 110 diffraction peaks (hexagonal axes) in the figure must completely overlap in the cubic case (220 for disordered NaCl structure and 440 for LiTiO$_2$ structure). Among the three compounds, the above result for NaPrTe$_2$ is inconsistent with the cubic LiTiO$_2$ structure reported previously,[20] as discussed in 3.3. In addition, there was no significant difference between the refinement results with the structural models without and with the vacancies or intersite defects. Therefore, the crystallographic parameters when the occupancy of each site is fixed to be 1 are listed in Table I. The volume of the unit cell at room temperature decreases in the order of $V$ = 399.569(5), 395.699(5), and 378.297(7) Å$^3$ for Ln = Pr, Nd, and Tb, respectively, which is consistent with the chemical trend of the ionic radii of the lanthanoids.

## 3.2 Physical Properties

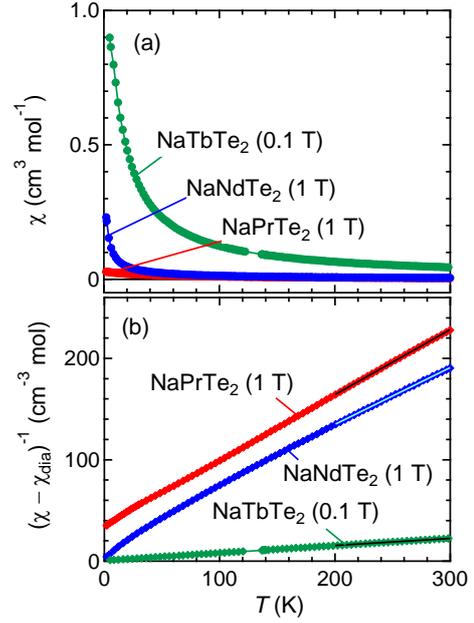

Figure 3. Temperature dependences of the magnetic susceptibility (a) and the inverse of magnetic susceptibility (b) of the polycrystalline samples of NaPrTe$_2$, NaNdTe$_2$, and NaTbTe$_2$. The c of NaPrTe$_2$ and NaNdTe$_2$ was measured in a magnetic field of $\mu_0 H$ = 1 T and that of NaTbTe$_2$ was measured in a magnetic field of 0.1 T. The data after subtraction of the diamagnetic contribution from core electrons ($\chi$ – $\chi_{dia}$)$^{-1}$, where $\chi_{dia}$ = –1.65 × 10$^{-4}$ cm$^3$ mol$^{-1}$ for NaPrTe$_2$ and NaNdTe$_2$ and $\chi_{dia}$ = –1.64 × 10$^{-4}$ cm$^3$ mol$^{-1}$ for NaTbTe$_2$, are shown in (b).[22] The solid lines in (b) show the result of a Curie-Weiss fit.

Figure 3 shows the temperature dependences of magnetic susceptibility and inverse magnetic susceptibility of the NaLnTe$_2$ (Ln = Pr, Nd, and Tb) polycrystalline samples, estimated assuming the samples are a single phase of NaLnTe$_2$. In the inverse magnetic susceptibility data, the diamagnetic contribution of the core electrons was subtracted.[22] At high temperatures, the ($\chi$ – $\chi_{dia}$)$^{-1}$ data for all samples exhibited a linear temperature dependence, following the Curie-Weiss law of $\chi$ – $\chi_{dia}$ = $C/(T - \theta_W)$, where $C$ and $\theta_W$ are the Curie constant and Weiss temperature, respectively. The Curie-Weiss fit to the ($\chi$ – $\chi_{dia}$)$^{-1}$ data between 200 and 300 K yields $C$ = 1.6, 1.8, and 14 cm$^3$ K mol-Ln$^{-1}$ and $\theta_W$ = –57, –40, and –17 K for Ln = Pr, Nd, and Tb, respectively. These $C$ values give effective moments of $p_{eff}$ = 3.5, 3.8, and 11 $\mu_B$/Ln, which are close to the effective moment of 4$f$ electrons of isolated Ln$^{3+}$ ions of $p_{eff}$ = $g_J\sqrt{J(J+1)}$ = 3.57, 3.62, and 9.72 $\mu_B$/Ln, respectively, where $g_J$ is the Landé $g$ factor. Although the samples contain impurity phases of Ln$_2$O$_2$Te and PrTe$_2$ with the volume fractions of at most 10 %, the agreement between the experimental and theoretical $p_{eff}$ values suggests that these three compounds are localized spin systems due to the 4$f$ electrons of the Ln$^{3+}$ ions. Note that the magnetic properties of Ln$_2$O$_2$Te have not been reported thus far[23] and PrTe$_2$ was reported to show Curie-Weiss magnetic susceptibility with the effective moment of 4.07 $\mu_B$/Pr.[24]

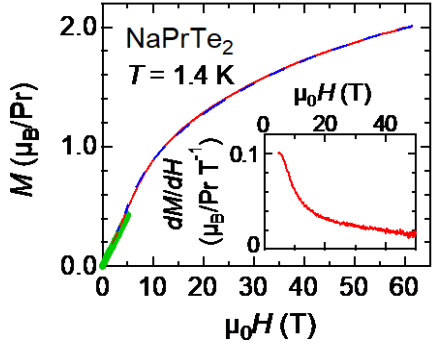

Figure 4. Magnetization curves of a powder sample of NaPrTe$_2$ measured up to 61 T at 1.4 K using a multilayered pulse magnet. The solid and broken curves indicate the data measured with increasing and decreasing magnetic fields, respectively. The filled plot indicates the data measured using a Magnetic Properties Measurement System at 1.8 K. The inset shows the $dM/dH$ of the magnetization curve measured with increasing magnetic fields shown in the main panel.

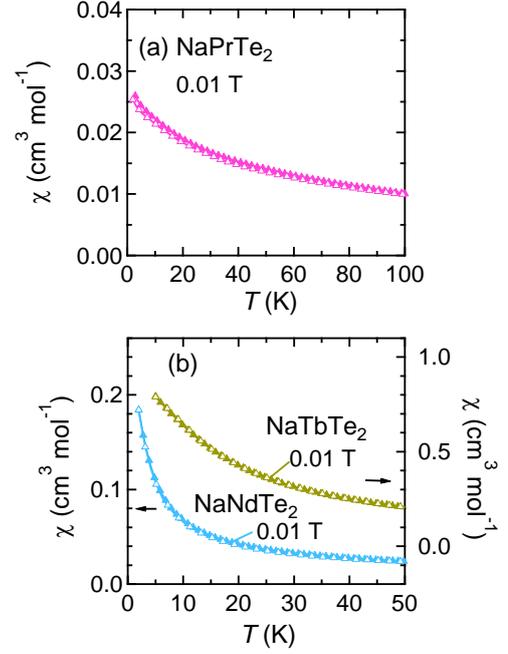

Figure 5. (a, b) Temperature dependence of field-cooled and zero-field-cooled magnetic susceptibility of (a) NaPrTe$_2$, (b) NaNdTe$_2$, and NaTbTe$_2$ polycrystalline samples measured at a magnetic field of $\mu_0 H = 0.01$ T. Filled and open plots indicate the zero-field-cooled and field cooled data, respectively. In (b), the left and right vertical axes indicate the data for NaNdTe$_2$ and NaTbTe$_2$, respectively.

The magnitudes of the magnetic interactions might be different between the three materials. As discussed above, the $\theta_W$ of NaNdTe$_2$ and NaTbTe$_2$ are –40 and –17 K, respectively, but the $(\chi - \chi_{dia})^{-1}$ data strongly deviate from linear behavior at low temperatures probably due to the crystalline field effect. The Curie-Weiss fits to the low-temperature data yielded negative Weiss temperatures of several K, meaning that the magnetic interactions at low temperatures are weak. In contrast, the $\theta_W$ of NaPrTe$_2$ determined by the Curie-Weiss fit to the 200-300 K data is –57 K and its $(\chi - \chi_{dia})^{-1}$ data decreased almost linearly toward the lowest measured temperature, possibly reflecting the moderately strong antiferromagnetic interaction between Pr$^{3+}$ spins even at low temperatures, unlike NaNdTe$_2$ and NaTbTe$_2$. The high-field magnetization data of NaPrTe$_2$ might be consistent with the presence of this antiferromagnetic interaction. As shown in Fig. 4, the magnetization of the NaPrTe$_2$ powder sample measured at 1.4 K continuously increased by applying a magnetic field, reaching 2.0 m$_B$/Pr at 61 T, which is 63% of the saturation magnetization $g_J J \mu_B = 3.2$ $\mu_B$ of an isolated Pr$^{3+}$ ion. The saturated magnetization is expected to be reached by applying a higher magnetic field. However, as seen in Fig. 3(b), the inverse magnetic susceptibility of NaPrTe$_2$ showed a small shoulder around 30 K and is smaller than the Curie-Weiss line at lower temperatures. This behavior can be realized by the presence of weakly coupled impurity spins. In contrast, it might also be due to the effect of crystalline field splitting. For NaPrTe$_2$, $J = 4$ magnetic moment of Pr$^{3+}$ in the local $D_{3d}$ symmetry splits into three singlet and three magnetic doublet ($2A_{1g} + A_{2g} + 3E_g$). In the latter case, the saturation magnetization will be a smaller value than $g_J J \mu_B = 3.2$ $\mu_B$, which is consistent with the observed $M = 2.0$ $\mu_B$/Pr at 61 T. To confirm the magnitudes of antiferromagnetic interaction and crystalline field effect at low temperatures, further experiments should be performed in the future studies.

The $\chi$ of NaLnTe$_2$ does not show an anomaly due to the magnetic transition. Figure 5 shows the field-cooled and zero-field-cooled magnetic susceptibility of NaLnTe$_2$ (Ln = Pr, Nd, and Tb) measured at a magnetic field of $\mu_0 H = 0.01$ T. The NaPrTe$_2$ data exhibited a small hysteresis probably due to a tiny amount of ferromagnetic or glassy impurity, but there is no anomaly corresponding to the magnetic order or spin glass formation. There is also no anomaly in the c data of NaNdTe$_2$ and NaTbTe$_2$ above 2 K. When a certain degree of antiferromagnetic interaction exists at low temperatures, as suggested in NaPrTe$_2$, the above results suggest that the geometrical frustration due to the triangular lattice works. However, when the magnetic interactions at low temperatures are weak, which is most likely in NaNdTe$_2$ and NaTbTe$_2$, the experiments at a lower temperature are required in order to reveal the magnetic ground states.

Figure 6 shows the electrical resistivity, r, and heat capacity divided by temperature, $C_p/T$, of NaPrTe$_2$. The electrical resistivity shown in Fig. 6(a) exhibits semiconducting behavior without an anomaly suggesting the presence of a phase transition. The $C_p/T$ data do not also exhibit a clear anomaly and is almost independent of the magnetic field. The concave downward behavior in the $C_p/T$ versus $T^2$ plot shown in the inset suggests the presence of a significant contribution of magnetic heat capacity. Rough estimation of the spin entropy below 20 K gives ~2 J K$^{-1}$ mol$^{-1}$, which is almost 30% of $R\ln 2$, where $R$ is the gas constant, although it is difficult to discuss the electronic state at low temperatures from this estimation.

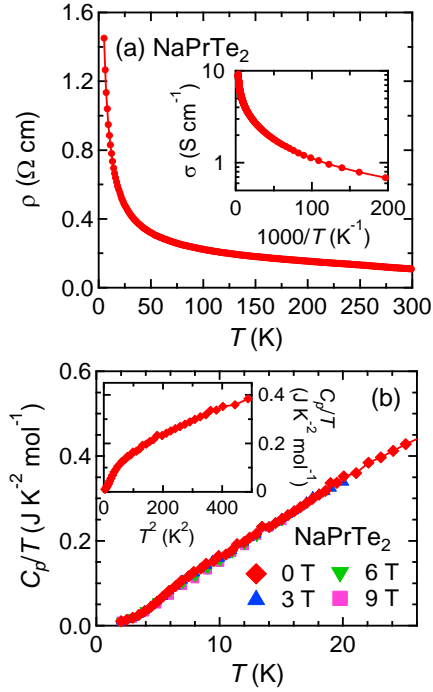

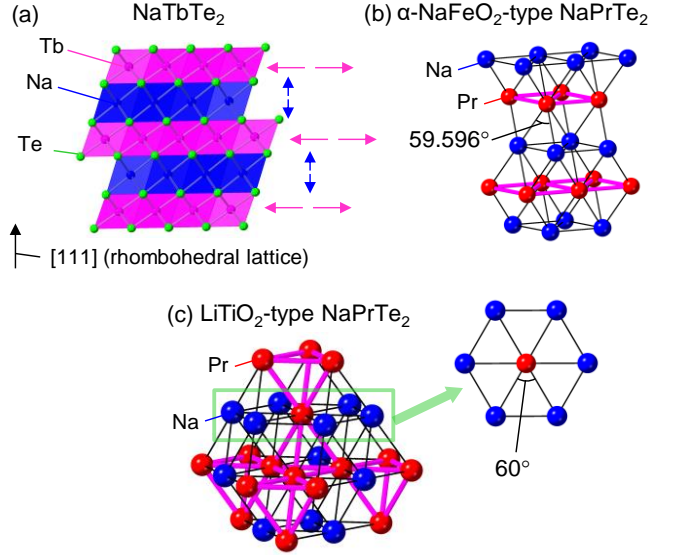

Figure 6. (a) Temperature dependence of electrical resistivity of a polycrystalline sample of NaPrTe$_2$. The inset shows an Arrhenius plot of electrical conductivity of NaPrTe$_2$. (b) Temperature dependence of heat capacity divided by temperature, $C_p/T$, of a polycrystalline sample of NaPrTe$_2$ measured at various magnetic fields. The inset shows the $C_p/T$ versus $T^2$ plot.

Figure 7. (a) Crystal structure of NaTbTe$_2$. TbTe$_6$ octahedra are compressed and NaTe$_6$ octahedra are elongated along [111] direction of the rhombohedral lattice, because a Tb atom is smaller than Na atom. (b, c) Na and Pr sublattices of α-NaFeO$_2$-type (b) and LiTiO$_2$-type NaPrTe$_2$. The thick solid lines show the triangular lattice and pyrochlore structure made of Pr$^{3+}$ atoms, respectively. In (b) and (c), the Na-Pr-Na angles are shown.

### 3.3 Polytypism in NaPrTe$_2$

We discuss the structural properties of NaLnTe$_2$, mainly focusing on the polytypism in NaPrTe$_2$. The NaPrTe$_2$ samples obtained in this study crystallize in the rhombohedral α-NaFeO$_2$ structure, in spite of the cubic LiTiO$_2$ structure in the previous study.[20] Both crystal structures are of the ordered NaCl type, where Na and Pr atoms form a slightly-distorted face centered cubic lattice, as shown in the lower panel of Fig. 1, but interconversion between α-NaFeO$_2$ and LiTiO$_2$ structures requires atomic diffusion. The lattice volumes for a formula unit are almost identical for both structures ($V/Z$ = 133.189(3) Å$^3$ for the α-NaFeO$_2$ type and $V/Z$ = 132.772(2) Å$^3$ for the LiTiO$_2$ type, where $Z$ is the number of the formula units in the unit cell), and the Pr-Te and Na-Te bond lengths, $d_{\text{Pr-Te}}$ and $d_{\text{Na-Te}}$, for both structures are also close to the same ($d_{\text{Pr-Te}}$ = 3.173 and 3.174 Å and $d_{\text{Na-Te}}$ = 3.262 and 3.254 Å for α-NaFeO$_2$ and LiTiO$_2$ types, respectively). Therefore, the difference in crystal structures between this and the previous studies is not due to the nonstoichiometry or lattice defects, but probably due to the different preparation methods, suggesting that NaPrTe$_2$ is a polytypic material.

The crystallographic parameters of NaPrTe$_2$ support the polytypic nature of this compound. In general, when an ABX$_2$ compound crystallizes in the ordered NaCl structure, the α-NaFeO$_2$ structure is selected in many cases. This is due to the high tolerance of this crystal structure, where the difference in ionic radii of the A and B atoms can be absorbed by the rhombohedral distortion of the X$_6$ octahedra. As seen for NaPrTe$_2$ in Fig. 1(a), each of the A (Na) and B (Pr) atoms form a triangular lattice. When the ionic radius of the A atom is larger than that of the B atom, the AX$_6$ octahedra are compressed and the BX$_6$ ones are stretched along [111] of the rhombohedral lattice, as shown in Fig. 7(a) using NaTbTe$_2$ as an example. It is natural that NaTbTe$_2$ crystallizes in the α-NaFeO$_2$ structure, because the sizes of the Na$^+$ and Tb$^{3+}$ ions are considerably different ($r_{\text{Na}^+}$ = 1.02 Å and $r_{\text{Tb}^{3+}}$ = 0.92 Å in octahedral coordination[25]).

This structural feature of NaPrTe$_2$ might be interesting from the dimensional control of the geometrically frustrated lattices, in which two different frustrated lattices with different dimensions can be realized as polytypes. As shown in the lower panel of Figs. 1(a,b), Pr$^{3+}$ ions form a two-dimensional triangular lattice in the rhombohedral α-NaFeO$_2$ structure, while they form a three-dimensional pyrochlore structure in the cubic LiTiO$_2$ structure. These two structures have various common points; they are geometrically frustrated lattices made of regular triangles and Pr$^{3+}$ ions in them have $-3m$ local symmetry and six nearest neighbors. Unfortunately, LiTiO$_2$-type NaPrTe$_2$ could not be obtained in this study, but NaPrTe$_2$ is expected to be a good platform for studying the effect of dimension of the lattice on geometrically frustrated magnetism.

## 4. Conclusion

Three tellurides, NaPrTe$_2$, NaNdTe$_2$, and NaTbTe$_2$ are found to be triangular lattice magnets with the α-NaFeO$_2$ structure, where lanthanoid atoms with $4f$ electrons form a triangular lattice. The α-NaFeO$_2$ structure is a new polymorph of NaPrTe$_2$, which has been reported to crystallize in the cubic LiTiO$_2$ structure. Structural parameters determined by the Rietveld analysis suggested that NaPrTe$_2$ is in the proximity of the phase boundary between the LiTiO$_2$ and α-NaFeO$_2$ types, most likely resulting in the polytypic nature of this compound. In these structural types, Pr atoms form the pyrochlore structure and triangular lattice, respectively, suggesting that NaPrTe$_2$ might be interesting from the perspective of the dimensional control of geometrically frustrated lattices. The magnetic susceptibility and heat capacity data indicate that NaPrTe$_2$ do not show long-range magnetic order or a spin-glass transition above 2 K.

**Acknowledgments**

The authors are grateful to D. Hirai, Z. Hiroi, and T. Yamauchi for their support in the heat capacity measurements. This work was partly carried out at the Materials Design and Characterization Laboratory and the International MegaGauss Laboratory under the Visiting Researcher Program of the Institute for Solid State Physics, University of Tokyo and supported by JSPS KAKENHI (Grant Nos.: 19H05823, 22H04467, and 23H01831). Synchrotron radiation experiments were conducted at the BL5S2 beamline of Aichi Synchrotron Radiation Center, Aichi Science and Technology Foundation, Aichi, Japan (Proposal Nos. 2020L4002 and 2021L2002). The high energy synchrotron powder X-ray diffraction experiments were conducted at the BL02B2 of SPring-8, Hyogo, Japan (Proposals No. 2020A1063).

1) R. Ballestracci and E. F. Bertaut, Bull. Soc. franç, Minér. Crist. **87**, 512-517 (1964).
2) R. Ballestracci, Bull. Soc. franç, Minér. Crist. **88**, 207-210 (1965).
3) M. Tromme, C. R. Acad. Sci. Sér. C **273**, 849-854 (1971).
4) R. Ballestracci and E. F. Bertaut, Bull. Soc. franç, Minér. Crist. **88**, 136-138 (1965).
5) T. Ohtani, H. Honjo, and H. Wada, Mat. Res. Bull. **22**, 829-840 (1987).
6) K. Stöwe, Z. anorg. allg. Chem. **623**, 1639-1643 (1997).
7) H. Masuda, T. Fujino, N. Sato, and K. Yamada, Mat. Res. Bull. **34**, 1291-1300 (1999).
8) A. K. Gray, B. R. Martin, and P. K. Dorhout, Z. Kristallogr. NCS **218**, 19 (2003).
9) L. J. Butts, N. Strickland, and B. R. Martin, Mat. Res. Bull. **44**, 854-859 (2009).
10) J. Fábry, L. Havlák, M. Kučeráková, and M. Dušek, Acta Cryst. **C70**, 533-535 (2014).
11) M. Brunel, F. De Bergevin, and M. Gondrand, J. Phys. Chem. Solids **33**, 1927-1940 (1972).
12) G. Blasse, J. inorg. nucl. Chem. **28**, 2444-2445 (1966).
13) Y. Hashimoto, M. Wakeshima, and Y. Hinatsu, J. Solid State Chem. **176**, 266-272 (2003).
14) M. M. Bordelon, E. Kenney, C. Liu, T. Hogan, L. Posthuma, M. Kavand, Y. Lyu, M. Sherwin, N. P. Butch, C. Brown, M. Graf, L. Balents, and S. D. Wilson, Nat. Phys. **15**, 1058-1064 (2019).
15) K. M. Ranjith, D. Dmytriieva, S. Khim, J. Sichelschmidt, S. Luther, D. Ehlers, H. Yasuoka, J. Wosnitza, A. A. Tsirlin, H. Kühne, and M. Baenitz, Phys. Rev. B **99**, 180401(1-7) (2019).
16) L. Ding, P. Manuel, S. Bachus, F. Gruβler, P. Gegenwart, J. Singleton, R. D. Johnson, H. C. Walker, D. T. Adroja, A. D. Hillier, A. A. Tsirlin, Phys. Rev. B **100**, 144432(1-7) (2019).
17) M. Baenitz, Ph. Schlender, J. Sichelschmidt, Y. A. Onykiienko, Z. Zangeneh, K. M. Ranjith, R. Sarkar, L. Hozoi, H. C. Walker, J.-C. Orain, H. Yasuoka, J. van den Brink, H. H. Klauss, D. S. Inosov, and Th. Doert, Phys. Rev. B **98**, 220409(1-5) (2018).
18) R. Sarkar, Ph. Schlender, V. Grinenko, E. Haeussler, P. J. Baker, Th. Doert, and H.-H. Klauss, Phys. Rev. B **100**, 241116(1-5) (2019).
19) K. M. Ranjith, S. Luther, T. Reimann, B. Schmidt, Ph. Schlender, J. Sichelschmidt, H. Yasuoka, A. M. Strydom, Y. Skourski, J. Wosnitza, H. Kühne, Th. Doert, and M. Baenitz, Phs. Rev. B **100**, 224417(1-11) (2019).
20) F. Lissener and T. Schleid, Z. Anorg. Allg. Chem. **629**, 1895-1897 (2003).
21) F. Izumi and K. Momma, Solid State Phenom. **130**, 15-20 (2007).
22) R. R. Gupta, in *Landolt-Börnstein*, *New Series*, *Group II* (Springer-Verlag, Berlin, 1986), Vol. 16, p. 402.
23) F. A. Weber and T. Schleid, Z. Anorg. Allg. Chem. **625**, 1833 (1999).
24) Y. S. Shin, C. W. Han, B. H. Min, H. J. Lee, C. H. Choi, Y. S. Kim, D. L. Kim, and Y. S. Kwon, Physica B **291**, 225 (2000).
25) R. D. Shannon, Acta Cryst. **A32**, 751 (1976).

*yokamoto@issp.u-tokyo.ac.jp
†Present address: Institute for Solid State Physics, The University of Tokyo, Kashiwa 277-8581, Japan